\documentclass[11pt,twoside]{gsp}
\usepackage{amsmath}
\usepackage{nccmath}
\input

\def\be{\begin{equation}} \def\ee{\end{equation}}
\def\ba   {\begin{array}}      \def\ea   {\end{array}}
\def\bean {\begin{eqnarray*}}  \def\eean {\end{eqnarray*}}
\def\bea{\begin{eqnarray}} \def\eea{\end{eqnarray}}

\def\pa {\partial}
\def\ti {\tilde}

\def\eq {\equiv}

\def\la{\lambda}

\newtheorem{rem}{Remark}
\newtheorem{theorem}{Theorem}[section]

\def\proof{\noindent {\bf Proof.\ }}
\def\qed{\vrule height0.6em width0.3em depth0pt}

\numberwithin{equation}{section}

\begin{document}
\thispagestyle{plain}
\title{Four Points Linearizable Lattice Schemes}


\author{Decio Levi and Christian Scimiterna}

\date{}

\maketitle
\comm{Communicated by XXX}\\
\begin{abstract}

We provide conditions for  a lattice scheme defined on a four points lattice to be linearizable by a point transformation. We apply the obtained conditions to a symmetry preserving difference scheme for the potential Burgers introduced by Dorodnitsyn \cite{db} and show that it is not linearizable.
\end{abstract}

\medskip

\noindent PACS numbers: 

\noindent Mathematics Subject Classification: 

\medskip

\setcounter{equation}{0}
\section{Introduction}
In a recent article \cite{ls} we extended to lattice equations the theorems introduced by Bluman and Kumei \cite{bk}  for proving the linearizability of nonlinear Partial Differential Equations (PDEs)  (for a recent extended review see \cite{bca}) based on the analysis of the symmetry properties of linear PDEs.  In \cite{ls}, following the analogy of the continuous case  we formulated a similar theory for linearizable Partial Difference Equations (P$\Delta$Es) on a fixed lattice. 

Here we extend the results of \cite{ls} to the case of a lattice scheme, i.e. when the lattice is not a priory given but it is defined by an equation so as to be able to perform a symmetry preserving discretization of the PDE.

In Section 2 we prove a theorem characterizing the symmetries of linear difference schemes on four lattice points  and in Section 3 we apply it to find conditions under which a  nonlinear difference scheme is linearizable.
These conditions are then applied to the symmetry preserving discretization of the potential Burgers. 
\setcounter{equation}{0}
\section{Symmetries of linear schemes}
In this Section we  define a difference scheme and  provide the symmetry conditions under which such a scheme is linearizable.  To do so in a definite way we limit ourselves to the case when the equation and the lattice are defined on four lattice points in the plane, i.e. we consider one scalar equation for a continuous
function of two (continuous) variables: $u_{m,n}=u(x_{m,n},t_{m,n})$ defined on four lattice points.
\begin{figure}[htbp] \label{fig1}
\begin{center}
\setlength{\unitlength}{0.1em}
\begin{picture}(200,140)(-50,-20)

  \put( 0,  0){\line(1,0){100}}
  \put( 100,  0){\line(-1,0){100}}

  \put( 0,100){\line(1,0){100}}
  \put( 100,100){\line(-1,0){100}}

  \put(  0, 0){\line(0,1){100}}
  \put(  0, 100){\line(0,-1){100}}

  \put(100, 0){\line(0,1){100}}
  \put(100, 100){\line(0,-1){100}}

   \put(97, -3){$\bullet$}
   \put(-3, -3){$\bullet$}
   \put(-3, 97){$\bullet$}
   \put(97, 97){$\bullet$}
  \put(-32,-13){$u_{m,n}$}
  \put(103,-13){$u_{m,n+1}$}
  \put(103,110){$u_{m+1,n+1}$}
  \put(-32,110){$u_{m+1,n,m}$}
\end{picture}
\caption{The  $\mathbb{Z}^2$ square-lattice where the equation  is defined}
\end{center}
\end{figure}
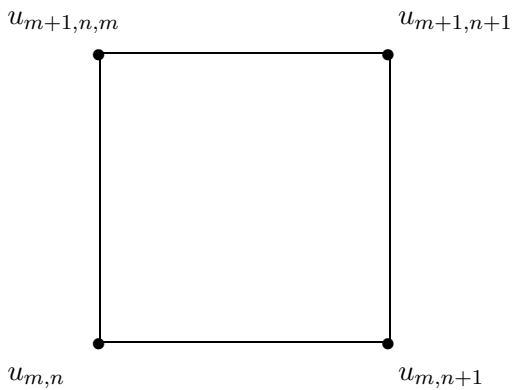
\subsection{The difference scheme}

As   we consider one scalar equation for a continuous
function of two (continuous) variables, a lattice will be a set of
points $P_i$, lying in the plane $\mathbb{R}^2$ and stretching in all directions
with, a priori,  no boundaries. The points $P_i$ in $\mathbb{R}^2$ will be labeled by two
discrete labels $P_{m,n}$. The Cartesian coordinates of the point $P_{m,n}$
will be $(x_{m,n}, t_{m,n})$ with $-\infty < m < \infty\ ,\ -\infty < n < \infty$. The value of the
dependent variable in the point $P_{m,n}$ will be denoted $u_{m,n} \equiv u(x_{m,n},t_{m,n})$.

A difference scheme will be a set of $b$ equations relating the values of
$\{x,t,u\}$ in a finite number of points. We start with one `reference point'
$P_{m,n}$ and define a finite number of points $P_{m+i,n+j}$ in the
neighborhood of $P_{m,n}$. They must lie on two different sets of curves,
two of which will be intersecting in $P_{m,n}$. Thus, the difference scheme will have the form
\be
\ba{c}
E_a\Big(\left\{ x_{m+i,n+j},t_{m+i,n+j},u_{m+i,n+j} \right\} \Big)=0
\ \ \ \ \ \ 1\le a \le b
\\*[2ex]
-i_1 \le i \le i_2\ \ \ \ -j_1 \le j \le j_2\ \ \ \
i_1,i_2,j_1,j_2 \in \mathbb{Z}^{\ge 0}.
\label{eq:2.1}
\ea
\ee

The situation is illustrated on Figure~2 in the case  of a present 7 points  lattice. Our convention is that $x$ increases as $m$ grows, $t$ increases as
$n$ grows (i.e. $x_{m+1,n}-x_{m,n} \eq h_1>0\ ,\ t_{m,n+1}-t_{m,n} \eq h_2>0$). The scheme on
Figure~2 could be used e.g. to approximate a differential equation of third
order in $x$, second in $t$. 

\bigskip
\begin{figure}[htbp] \label{fig2}
\begin{center}
\setlength{\unitlength}{0.1em}
\begin{picture}(160,160)
\put(0,0){\vector(1,0){160}} \put(0,0){\vector(0,1){160}}

\put(150,-5){\makebox(0,0){$x$}} \put(-5,150){\makebox(0,0){$t$}} \qbezier(15,15)(32,52)(55,85)
\qbezier(55,85)(90,132)(145,150) \qbezier(15,150)(29,148)(55,85) \qbezier(55,85)(80,38)(145,30)
\qbezier(45,165)(62,150)(82,114) \qbezier(82,114)(100,74)(130,76)\qbezier(10,120)(28,140)(85,160)

\put(55,85){\makebox(0,0){$\bullet$}} \put(60,85){\makebox(0,0)[l]{$P_{m,n}$}}

\put(25,35){\makebox(0,0){$\bullet$}} \put(28,35){\makebox(0,0)[l]{$P_{m-1,n}$}}

\put(82,114){\makebox(0,0){$\bullet$}} \put(89,114){\makebox(0,0)[l]{$P_{m+1,n}$}}

\put(60,150){\makebox(0,0){$\bullet$}} \put(65,147){\makebox(0,0)[l]{$P_{m+1,n+1}$}}

\put(120,140){\makebox(0,0){$\bullet$}} \put(125,135){\makebox(0,0)[l]{$P_{m+2,n}$}}

\put(30,135){\makebox(0,0){$\bullet$}} \put(34,133){\makebox(0,0)[l]{$P_{m,n+1}$}}

\put(105,40){\makebox(0,0){$\bullet$}} \put(105,45){\makebox(0,0)[l]{$P_{m,n-1}$}}
\end{picture}
\caption{Points on a two dimensional lattice}
\end{center}
\end{figure}
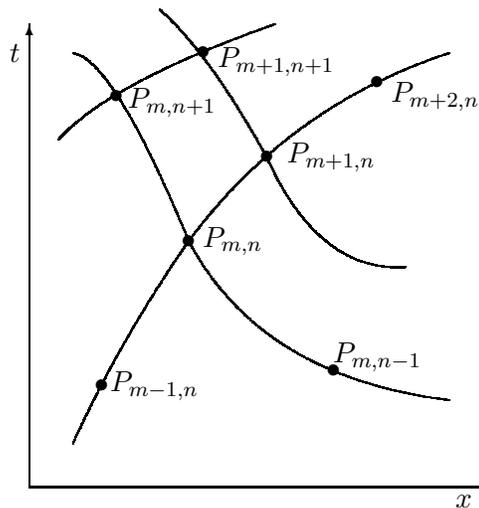

The value of $b$, the maximum number of different equations we consider, depends on the kind of problems we are considering.  Starting
from the reference point $P_{m,n}$ and a given number of neighboring points, it
must be possible to calculate the values of $\{x,t,u\}$ in all points in a unique way. This
requires a minimum of three equations to calculate the independent variables $(x,t)$ in
two directions and the dependent variable $u$ in all points. 
With one dependent variable in $\mathbb R^2$, at most we can set $b=5$. Of the five equations in (\ref{eq:2.1}), four determine completely the lattice, one
the difference equation.  If we choose $b=3$ than two define the lattice and one the difference equation and we are solving an initial value problem when both the equation and the lattice are defined from given initial conditions. If a continuous limit exists, \eqref{eq:2.1} represent a PDE in two variables. The equations determining the
lattice will reduce to identities (like $0=0$).

As an example of difference scheme, let us consider the simplest and most
standard lattice, namely a uniformly spaced orthogonal lattice and a difference
equation approximating the linear heat equation on this lattice. The 5 equations
(\ref{eq:2.1}) in this case are:
\bea
x_{m+1,n}-x_{m,n}=h_1\ \ \ \ \ \ t_{m+1,n}-t_{m,n}=0
\label{eq:2.3a}
\eea
\bea
x_{m,n+1}-x_{m,n}=0\ \ \ \ \ \ t_{m,n+1}-t_{m,n}=h_2
\label{eq:2.3b}
\eea
\bea
\frac{u_{m,n+1}-u_{m,n}}{h_2}=\frac{u_{m+1,n}-2u_{m,n}+u_{m-1,n}}{(h_1)^2}
\label{eq:2.4}
\eea
where $h_1$ and $h_2$ are two constants.
\begin{figure} \label{fig3}
\setlength{\unitlength}{1mm}
\begin{center}
\begin{picture}(100,51)
 \linethickness{1.pt}
     \put(80,10) {\circle*{1,5}}
  \put(50,10) {\circle*{1,5}}
   \put(50,40) {\circle*{1,5}}
 \put(20,10) {\circle*{1,5}}
\put(49,07){$u_{m,n}$}
\put(79,07){$u_{m+1,n}$}
\put(19,07){$u_{m-1,n}$}
\put(51,39){\boldmath $u_{m,n+1}$}
\put(40,50){\line(1,-1){55}}
\put(10,10) {\line(1,0){85}}
\put(50,10) {\line(0,1){40}}
\put(60,50){\line(-1,-1){55}}

\end{picture}
\end{center}
\caption[]{Four points  in the case of the heat equation. }
\end{figure}
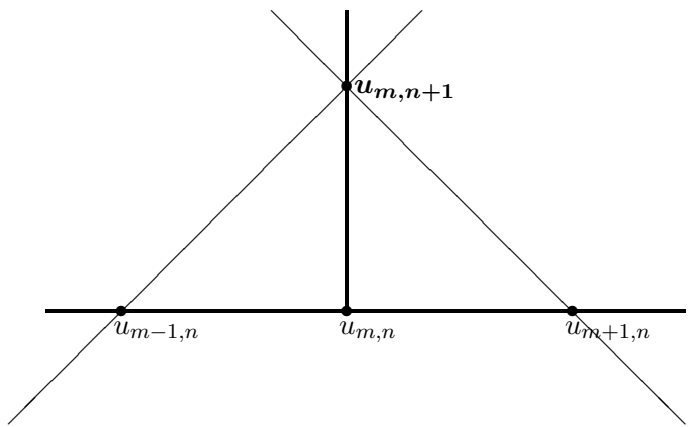

The example is simple;  the lattice equations can be solved
explicitly to give
\bea
x_{m,n}=h_1 m + x_0\ \ \ \ t_{m,n}=h_2 n + t_0.
\label{eq:2.5}
\eea
The usual choice is $x_0=t_0=0\ ,\ h_1=h_2=1$ and then $x$ is simply identified
with $m$, $t$ with $n$. 
The example suffices to bring out several points:
\begin{enumerate}
\item Four equations are needed to describe completely the lattice but in this case there is a compatibility condition. In the whole generality two equations are sufficient and provide the lattice starting from some initial conditions.
\item Four points are needed for equations of second order in $x$, first in $t$.
Only three figure in the lattice equation, namely $P_{m+1,n}, P_{m,n}$ and
$P_{m,n+1}$. To get the fourth point, $P_{m-1,n}$, we shift $m$ down by one unit
the equations (\ref{eq:2.3a}-\ref{eq:2.4}).
\item An independence condition  is needed to be able to solve
for $x_{m+1,n}$, $t_{m+1,n}$, $x_{m,n+1}$, $t_{m,n+1}$ and $u_{m,n+1}$.
\end{enumerate}
We need the more complicated two index notation to
describe arbitrary lattices and to formulate the symmetry algorithm.
\subsection{Symmetries of the difference scheme}

We are interested in point transformations of the type
\be
\ti{x}=F_{\la}(x,t,u)\ \ \ \ \ti{t}=G_{\la}(x,t,u)\ \ \ \ \ti{u}=H_{\la}(x,t,u)
\label{eq:2.6}
\ee
where $\la$ is a group parameter, such that when $(x,t,u)$ satisfy the system
(\ref{eq:2.1}) then $(\ti x,\ti t, \ti u)$ satisfy the same system. The
transformation acts on the entire space $(x,t,u)$, at least locally, i.e. in
some neighborhood of the reference point $P_{m,n}$, including all points
$P_{m+i,n+j}$ figuring in equation (\ref{eq:2.1}). That means that the same
functions $F,G$ and $H$ determine the transformation of all points. The
transformations (\ref{eq:2.6}) are generated by the vector field
\be
\hat X=\xi(x,t,u)\pa_x + \tau(x,t,u)\pa_t + \phi(x,t,u)\pa_u.
\label{eq:2.7}
\ee

The symmetry algebra of the system (\ref{eq:2.1}) is the
Lie algebra of the local symmetry group of local point transformations. An infinitesimal symmetry \eqref{eq:2.7} is a symmetry of (\ref{eq:2.1}) if (\ref{eq:2.1}) is invariant under a transformation  \eqref{eq:2.6}. To check it 
 we must prolong the action of the vector field $\hat X$ from the reference
point $(x_{m,n},t_{m,n},u_{m,n})$ to all points figuring in the system (\ref{eq:2.1}).
Since the transformations are given by the same functions $F,G$ and $H$ at all
points, the prolongation of the vector field (\ref{eq:2.7}) is obtained simply
by evaluating the functions $\xi, \tau$ and $\phi$ at the corresponding points.
In other words, we have
\be
\ba{l}
pr\, \hat X= \sum_{m,n} \Big[ \xi(x_{m,n},t_{m,n},u_{m,n})\pa_{x_{m,n}} +
\tau(x_{m,n},t_{m,n},u_{m,n})\pa_{t_{m,n}}
\\*[2ex]
 \ \ \ \ \ \ \ \ \ \ \ \ \ \ \ \ \ \
+\phi(x_{m,n},t_{m,n},u_{m,n})\pa_{u_{m,n}}\Big],
\ea
\label{eq:2.8}
\ee
where the summation is over all points figuring in the system (\ref{eq:2.1}).
The invariance requirement is formulated in terms of the prolonged vector field
as
\be
\mbox{pr} \hat X\, E_a\left|_{E_{c}=0}=0\ \ \ \ 1 \le a,c \le b. \right.
\label{eq:2.9}
\ee

Just as in the case of PDE's \cite{olver}, we can turn equation
(\ref{eq:2.9}) into an algorithm for determining the symmetries, i.e. finding the
coefficients in vector field (\ref{eq:2.7}) \cite{lw}.

\subsection{Symmetries of a linear partial difference scheme}
To be able to linearize a difference scheme using the knowledge of its symmetries we must be able to characterize the symmetries of a linear scheme. To do so in this subsection we prove a theorem on the structure of the symmetries of a linear partial difference scheme:
\begin{theorem} \label{pippo}
Necessary and sufficient conditions for three difference equations $\mathfrak E_{m,n}=0$, $\mathfrak F_{m,n}=0$ and $\mathfrak G_{m,n}=0$ defined on four points $\{(m,n),(m+1,n),(m,n+1),(m+1,n+1)\}$ for a scalar function $u_{m,n}(x_{m,n}, t_{m,n})$ and the lattice variables $x_{m,n}$ and $t_{m,n}$ to be linear is that they are invariant with respect to the following infinitesimal generator
\bea \label{2.3.1}
\hat X_{m,n} = v_{m,n} \partial_{u_{m,n}} + \chi_{m,n} \partial_{x_{m,n}} + \eta_{m,n} \partial_{t_{m,n}},
\eea
where the discrete functions $v_{m,n}$, $\chi_{m,n}$, $\eta_{m,n}$ satisfy three linear equations \newline $v_{m+1,n+1}=\mathfrak e_{m,n}$, $\chi_{m+1,n+1}=\mathfrak f_{m,n}$ and $\tau_{m+1,n+1}=\mathfrak g_{m,n}$. The functions $\mathfrak e$, $\mathfrak f$ and $\mathfrak g$ depend just on the functions  ($v_{m,n}$, $\chi_{m,n}$, $\eta_{m,n}$) in the  points $(m,n)$, $(m+1,n)$ and $(m,n+1)$  and are given by:
\begin{align} \nonumber
\mathfrak e_{0,0}&=a_1 v_{0,0} + a_2 v_{0,1} + a_3 v_{1,0}  + a_4 \chi_{0,0} + a_5 \chi_{0,1} + a_6 \chi_{1,0} + a_7 \eta_{0,0}+\\ \nonumber & +a_8 \eta_{0,1} + a_9 \eta_{1,0},\\ \nonumber
\mathfrak f_{0,0}&=b_1 v_{0,0} + b_2 v_{0,1} + b_3 v_{1,0}  + b_4 \chi_{0,0} + b_5 \chi_{0,1} + b_6 \chi_{1,0} + b_7 \eta_{0,0}+\\ \nonumber &+b_8 \eta_{0,1} + b_9 \eta_{1,0},\\   \label{2.3.8}
\mathfrak g_{0,0}&=c_1 v_{0,0} + c_2 v_{0,1} + c_3 v_{1,0}  + c_4 \chi_{0,0} + c_5 \chi_{0,1} + c_6 \chi_{1,0} + c_7 \eta_{0,0}+\\ \nonumber &+c_8 \eta_{0,1} + c_9 \eta_{1,0}, 
\end{align}
where $a_1$, $\cdots$, $c_9$ depend only on the lattice indices and where, here and in the following, for the sake of simplicity we  set  in any discrete variable on the square $z_{m+i,n+j}=z_{i,j}$. The linear equations $\mathfrak E_{m,n}=0$, $\mathfrak F_{m,n}=0$ and $\mathfrak G_{m,n}=0$ have the form:
\begin{align} \nonumber
u_{1,1}&=a_1 u_{0,0}+a_2 u_{0,1}+a_3 u_{1,0} + a_4 x_{0,0}+a_5 x_{0,1}+a_6 x_{1,0}+a_7 t_{0,0}+\\ \nonumber &+a_8 t_{0,1}+a_9 t_{1,0}, \\ \nonumber
x_{1,1}&=b_1 u_{0,0}+b_2 u_{0,1}+b_3 u_{1,0} + b_4 x_{0,0}+b_5 x_{0,1}+b_6 x_{1,0}+b_7 t_{0,0}+\\ \nonumber &+b_8 t_{0,1}+b_9 t_{1,0}, \\  \label{2.3.7}
t_{1,1}&=c_1 u_{0,0}+c_2 u_{0,1}+c_3 u_{1,0} + c_4 x_{0,0}+c_5 x_{0,1}+c_6 x_{1,0}+c_7 t_{0,0}+\\ \nonumber &+c_8 t_{0,1}+c_9 t_{1,0}.
\end{align}
\end{theorem}
\proof To prove this Theorem we require that  a generic P$\Delta$E $F_{m,n}=0$, depending on a scalar function $u_{m,n}(x_{m,n}, t_{m,n})$ and the lattice variables $x_{m,n}$ and $t_{m,n}$ in the four points $\{(m,n),(m+1,n),(m,n+1),(m+1,n+1)\}$, i.e. 12 variables,  be invariant under the prolongation of \eqref{2.3.1}, as given by \eqref{eq:2.8}. 
The invariance condition \eqref{eq:2.9}, when $\xi_{m,n}(x,t,u)=\chi_{m,n}$, $\tau_{m,n}(x,t,u)=\eta_{m,n}$ and $\phi_{m,n}(x,t,u)=v_{m,n}$ implies that $F_{m,n}$ should depend on  a set of 11 invariants $v_{m,n}$, $\chi_{m,n}$ and $\eta_{m,n}$ independent
\begin{align} \nonumber
&L_1=v_{0,0} u_{0,1}-v_{0,1} u_{0,0}, \quad  &L_2=v_{0,0} u_{1,0}-v_{1,0} u_{0,0}, \\ \nonumber &L_3=v_{0,0} u_{1,1}-\mathfrak e_{0,0}u_{0,0},  &L_4=v_{0,0} x_{0,1}-\chi_{0,1} u_{0,0}, \\ \nonumber
&L_5=v_{0,0} x_{1,0}-\chi_{1,0} u_{0,0}, \quad  &L_6=v_{0,0} x_{1,1}-\mathfrak f_{0,0}u_{0,0}, \\ \nonumber &L_7=v_{0,0} t_{0,1}-\eta_{0,1} u_{0,0}, \quad &L_8=v_{0,0} t_{1,0}-\eta_{1,0} u_{0,0}, \\ \nonumber
&L_9=v_{0,0} t_{1,1}-\mathfrak g_{0,0} u_{0,0}, \quad &L_{10}=v_{0,0} x_{0,0} - \chi_{0,0} u_{0,0}, \\ &\qquad \qquad \qquad  L_{11}=v_{0,0} t_{0,0} - \eta_{0,0} u_{0,0}.
\label{2.3.6}
\end{align}

 As $F_{m,n}$ should not depend on the functions  ($v_{m,n}$, $\chi_{m,n}$, $\eta_{m,n}$) in the  points $(m,n)$, $(m+1,n)$ and $(m,n+1)$ we have nine constraints given by $\frac{\partial F_{m,n}}{\partial v_{m+i,n+j}}=0$, $\frac{\partial F_{m,n}}{\partial \chi_{m+i,n+j}}=0$ and $\frac{\partial F_{m,n}}{\partial tau_{m+i,n+j}}=0$ with $(i,j)=(0,0),(0,1),(1,0)$. These are first order partial differential equations for the function $F_{m,n}$ with respect to the 11 invariants which we can solve on the characteristics to define three invariants
\begin{align} \nonumber 
K_1&=v_{0,0} \{ u_{1,1}-[\mathfrak e_{0,0,v_{0,1}}u_{0,1}+\mathfrak e_{0,0,v_{1,0}}u_{1,0}+\mathfrak e_{0,0,\chi_{0,0}}x_{0,0}+\mathfrak e_{0,0,\chi_{0,1}}x_{0,1} +\\ \nonumber &+\mathfrak e_{0,0,\chi_{1,0}}x_{1,0} +\mathfrak e_{0,0,\eta_{0,0}}t_{0,0}+\mathfrak e_{0,0,\eta_{0,1}}t_{0,1}+\mathfrak e_{0,0,\eta_{1,0}}t_{1,0}] \} - u_{0,0} \{ \mathfrak e_{0,0}- \\ \nonumber &-[\mathfrak e_{0,0,v_{0,1}}v_{0,1}+\mathfrak e_{0,0,v_{1,0}}v_{1,0}+ \mathfrak e_{0,0,\chi_{0,0}}\chi_{0,0}+\mathfrak e_{0,0,\chi_{0,1}}\chi_{0,1}+\mathfrak e_{0,0,\chi_{1,0}}\chi_{1,0}+\\ \nonumber &+\mathfrak e_{0,0,\eta_{0,0}}\eta_{0,0}+\mathfrak e_{0,0,\eta_{0,1}}\eta_{0,1}+\mathfrak e_{0,0,\eta_{1,0}}\eta_{1,0}] \}, \\ \nonumber 
K_2&=v_{0,0} \{ u_{1,1}-[\mathfrak f_{0,0,v_{0,1}}u_{0,1}+\mathfrak f_{0,0,v_{1,0}}u_{1,0}+\mathfrak f_{0,0,\chi_{0,0}}x_{0,0}+\mathfrak f_{0,0,\chi_{0,1}}x_{0,1}+\\ \nonumber &+\mathfrak f_{0,0,\chi_{1,0}}x_{1,0} +\mathfrak f_{0,0,\eta_{0,0}}t_{0,0}+\mathfrak f_{0,0,\eta_{0,1}}t_{0,1}+\mathfrak f_{0,0,\eta_{1,0}}t_{1,0}] \} - u_{0,0} \{ \mathfrak f_{0,0}-\\ \nonumber &-[\mathfrak f_{0,0,v_{0,1}}v_{0,1}+\mathfrak f_{0,0,v_{1,0}}v_{1,0}+\mathfrak f_{0,0,\chi_{0,0}}\chi_{0,0}+\mathfrak f_{0,0,\chi_{0,1}}\chi_{0,1}+\mathfrak f_{0,0,\chi_{1,0}}\chi_{1,0}+\\ \nonumber &+\mathfrak f_{0,0,\eta_{0,0}}\eta_{0,0}+\mathfrak f_{0,0,\eta_{0,1}}\eta_{0,1}+\mathfrak f_{0,0,\eta_{1,0}}\eta_{1,0}] \}, \\ \nonumber 
K_3&=v_{0,0} \{ u_{1,1}-[\mathfrak g_{0,0,v_{0,1}}u_{0,1}+\mathfrak g_{0,0,v_{1,0}}u_{1,0}+\mathfrak g_{0,0,\chi_{0,0}}x_{0,0}+\mathfrak g_{0,0,\chi_{0,1}}x_{0,1}+\\ \nonumber &+\mathfrak g_{0,0,\chi_{1,0}}x_{1,0} +\mathfrak g_{0,0,\eta_{0,0}}t_{0,0}+\mathfrak g_{0,0,\eta_{0,1}}t_{0,1}+\mathfrak g_{0,0,\eta_{1,0}}t_{1,0}] \} - u_{0,0} \{ \mathfrak g_{0,0}-\\ \nonumber &-[\mathfrak g_{0,0,v_{0,1}}v_{0,1}+\mathfrak g_{0,0,v_{1,0}}v_{1,0}+ \mathfrak g_{0,0,\chi_{0,0}}\chi_{0,0}+\mathfrak g_{0,0,\chi_{0,1}}\chi_{0,1}+\mathfrak g_{0,0,\chi_{1,0}}\chi_{1,0}+\\ \label{2.3.4} &+\mathfrak g_{0,0,\eta_{0,0}}\eta_{0,0}+\mathfrak g_{0,0,\eta_{0,1}}\eta_{0,1}+\mathfrak g_{0,0,\eta_{1,0}}\eta_{1,0}] \}.
\end{align}
 By construction the three invariants $K_i$, $i=1,2,3$ are independent and the three equations  $\mathfrak E_{m,n}=0$, $\mathfrak F_{m,n}=0$ and $\mathfrak G_{m,n}=0$ must be defined in terms of them. The three invariants $K_3$, $L_3$ and $M_3$  still depend on the functions  ($v_{m,n}$, $\chi_{m,n}$, $\eta_{m,n}$) in the  points $(m,n)$, $(m+1,n)$ and $(m,n+1)$ while they should depend just on the variables ($u_{m,n}$, $x_{m,n}$, $t_{m,n}$) in the  points $(m,n)$, $(m+1,n)$, $(m,n+1)$ and $(m+1,n+1)$. The derivatives   $F_{m,n,K_i}$, $i=1,2,3$ will  satisfy a set of nine linear equations whose coefficients will form a matrix $\mathfrak A$ 9x3. The matrix $\mathfrak A$ can have rank 3, 2 or 1. In the case of rank 3 we have $F_{m,n,K_i}=0$, $i=1,2,3$ i.e. the function $F_{m,n}$ does not depend on the 3 invariants. If the rank of $\mathfrak A$ is 2 or 1 we can have at most two independent invariants. If we want to have three invariants we need to require that the coefficients of the matrix $\mathfrak A$ be zero, i.e. defining $\alpha_1=v_{0,0}$, $\alpha_2=v_{0,1}$, $\alpha_3=v_{1,0}$, $\cdots$, $\alpha_9=\eta_{1,0}$ we have $\frac{\partial K_p}{\partial \alpha_q}=0$ $p=1,2,3$, $q=1,\cdots,9$. The equations $\frac{\partial K_p}{\partial \alpha_q}=0$ are linear homogeneous expressions in $u_{i,j}$, $x_{i,j}$ and $t_{i,j}$ with coefficients depending on $v_{i,j}$, $\chi_{i,j}$ and $\eta_{i,j}$, for appropriate values of $i$ and $j$.  Consequently \eqref{2.3.7}.
Than  $\frac{\partial K_p}{\partial \alpha_q}=0$ turn out to be a set of 159 overdetermined partial differential equations for  the functions $\mathfrak e_{m,n}$, $\mathfrak f_{m,n}$ and $\mathfrak g_{m,n}$ whose solution \eqref{2.3.8} is obtained using Maple. It depends on 27 integration constants which must be set equal zero if \eqref{2.3.7} does not depend on $v_{i,j}$, $\chi_{i,j}$ and $\eta_{i,j}$.
\qed

A few remarks can be derived from Theorem \ref{pippo} and must be stressed. 

\begin{rem} \label{r1}
The equation for $u_{m,n}$ and those for the lattice variables $x_{m,n}$ and $t_{m,n}$ are independent, however the functions appearing in the symmetry \eqref{2.3.1} do not satisfy equations independent from those satisfied by the lattice scheme. In fact these symmetries correspond to independent  superposition laws for the equation and the lattice. 
\end{rem}

\begin{rem} \label{r2}
If the linear equation for $u_{m,n}$ is autonomous than the coefficients $\{ a_4, \cdots, a_9 \}$ are zero. The variable  $v_{m,n}$ will satisfy a similar equation but the lattice equations can depend linearly on  $u_{m,n}$. 
\end{rem}

\begin{rem} \label{r3}
The proof of Theorem \ref{pippo} does not depends on the position of the four lattice points considered, i.e. $\{(m,n),(m+1,n),(m,n+1),(m+1,n+1)\}$. The same result is also valid if the four points are put on the triangle shown in Fig. 3, i.e. $\{(m,n),(m+1,n),(m-1,n),(m,n+1)\}$.
\end{rem}
\setcounter{equation}{0}
\section{Linearizable nonlinear schemes}
In this article each equation of a difference scheme is an equation for the continuous  variable $u_{m,n}$, $x_{m,n}$ and $t_{m,n}$. If the equations for the lattice variables, $x_{m,n}$ and $t_{m,n}$, are solvable we get
\bea \label{f1}
x_{m,n} = \mathcal X(m,n,c_0,c_1,\cdots), \qquad t_{m,n} = \mathcal T(m,n,d_0,d_1,\cdots),
\eea 
and then the remaining equation for the  variable $u_{m,n}$ depends explicitly on $n$ and $m$, on the integration constants contained in (\ref{f1}) and turns out to be an algebraic, maybe transcendental, equation of $u_{m,n}$ in the various lattice points involved in the equation. So the difference scheme reduce to a non autonomous  equation on a fixed lattice and for its linearization we can apply the results of \cite{ls}.
 
 If the equations for the lattice are not solvable the difference scheme can be thought as a system of coupled equations for the  variables $u_{m,n}$,  $x_{m,n}$ and $t_{m,n}$ on a fixed lattice. In this way we can apply to  the equations of the scheme the results of \cite{ls} and, taking into account the results of the previous section, we can propose the following linearizability theorem:
\begin{theorem} \label{tt1}
A nonlinear difference scheme (\ref{eq:2.1}) involving $i_1+i_2$ different points in the $m$ index and $j_1+j_2$ in the $n$ index for 
 a scalar function $u_{m,n}$ of a $2$--dimensional space of coordinates  $x_{m,n}$ and $t_{m,n}$ will be linearizable by a point transformation 
\bea \label{t2} 
w_{m,n}(y_{m,n},z_{m,n})&=&f(x_{m,n},t_{m,n},u_{m,n}),\,  y_{m,n}=g(x_{m,n},t_{m,n},u_{m,n}), \\ \nonumber && \quad z_{m,n}=k(x_{m,n},t_{m,n},u_{m,n}),
\eea
 to a linear difference scheme of the kind of (\ref{2.3.7}) for $w_{m,n}$, $y_{m,n}$ and $z_{m,n}$  if it possesses a symmetry generator 
 \bea \label{t3}
 \hat X &=&  \xi(x,t,u) \partial_{x} + \phi(x,t,u) \partial_t+\psi(x,t,u) \partial_u, \\ \nonumber \xi(x,u)&=&\alpha(x,t,u) y, \quad \phi(x,t,u)=\beta(x,t,u) z, \quad \psi(x,t,u)=\gamma(x,t,u) w
 \eea
  with $\alpha$, $\beta$ and $\gamma$  given functions of their arguments and $y$, $z$ and $w$  an arbitrary solution of (\ref{2.3.8}). 
\end{theorem}  
In the following we will consider the application of this theorem to a difference scheme which one would hope that it is linearizable as is a symmetry preserving discretization of a linearizable PDE, the potential Burgers equation \cite{olver}.
\subsection{Application}
We consider here the discretization of the potential Burgers presented by Dorodnitsyn et. al. \cite{db} and show that, even if it is reducible by a point transformation to the discrete scheme of the heat equation, is not linearizable by a point transformation. As a consequence we have that also the symmetry preserving discretization of the heat equation presented by Dorodnitsyn et. al. is not a linear difference scheme.

The symmetry preserving discretization of the potential Burgers is given by the following scheme
\bea \label{e1}
&\frac{\Delta x}{\tau} &= \frac{1}{h^+ + h^-} \Big [\frac{h^-}{h^+}(w_+-w)+\frac{h^+}{h^-}(w-w_-) \Big ] \\ \label{e1a}
& e^{\hat w - w -\frac{\Delta^2 x}{2 \tau}} &= 1+\frac{2\tau}{(h^+)^2} \Big [ \frac{w_+-w}{h^+}-\frac{w-w_-}{h^-} \Big ]\\
 \label{e1bb}
&\tau=t_{m,n+1}-t_{m,n},& \qquad t_{m+1,n}=t_{m-1,n}=t_{m,n}=t.
\eea
where $\tau$ is a constant and 
$$ w=w_{m,n}(x_{m,n},t_{m,n}), \, \hat w=w_{m,n+1}, \, w_-=w_{m-1,n}, \, w_+=w_{m+1,n}, $$  $$\Delta x=x_{m,n+1}-x_{m,n}, \quad h^+=x_{m+1,n}-x_{m,n}, \quad h^-=x_{m,n}-x_{m-1,n}.$$
Eqs. (\ref{e1}, \ref{e1a}) are written in terms of the discrete invariants $\mathcal I_2, \, \mathcal I_3, \, \mathcal I_4$ on the stencil defined in terms of ($\tau, \, x,\, \Delta x, \, h^+, \,  h^-, \, w, \, \hat w, \, w_+, \, w_-$) of the finite point symmetries of continuous potential Burgers equation 
\bea \label{e1c}
w_t=w_{xx}- \frac{1}{2} w_x^2,
\eea 
\bea \label{e2}
&\hat X_1 = \partial_t, \quad \hat X_2 = \partial_x, \quad &\hat X_3 = t \partial_x + x \partial_t, \quad \hat X_4 = 2 t \partial_t + x \partial_x, \\ \nonumber
&\hat X_5 = \partial_w, \quad &\hat X_6 = t^2 \partial_t + tx \partial_x + \Big ( \frac{1}{2} x^2 + t \Big ) \partial_w,
\eea 
\bea \label{e3}
\mathcal I_1 &=& \frac{H^+}{h^-}, \quad \mathcal I_2 = \frac{\tau^{1/2}}{h^+} e^{\frac{1}{2}(w-\hat w) +\frac{\Delta^2 x}{4 \tau}} ,\\ \nonumber
 \mathcal I_3 &=& \frac{1}{4}\frac{h^{+2}}{\tau} + \frac{h^{+2}}{h^+ + h^-} \Big [ \frac{w_+ - w}{h^+} - \frac{w - w_-}{h^-} \Big ], \\ \nonumber
 \mathcal I_4&=& \Delta x \frac{h^+}{\tau} -\frac{2 h^+}{h^+ + h^-} \Big [\frac{h^-}{h^+} ( w_+ - w) + \frac{h^-}{h^+} ( w - w_-) \Big ], 
\eea
and goes into it in the continuos limit.

Eqs. (\ref{e1}, \ref{e1a}) are related to a symmetry preserving discretization of the heat equation for $u_{m,n}$ by the point transformation $w_{m,n}=-2 \log(u_{m,n})$.  However it is not completely obvious if (\ref{e1}, \ref{e1a}) are reducible to a linear discrete equation i.e. if it possess, as its continuos counterpart (\ref{e1c}), an infinite dimensional symmetry $\hat X =u(x,t) e^{-w}  \partial_w$ with $u(x,t)$ solution of the linear heat equation $u_t = u_{xx}$.  

We can apply on the lattice scheme (\ref{e1}, \ref{e1a}, \ref{e1bb}) the symmetry generator 
\bea \label{e4}
\hat X = \psi(x,t,w) u \partial_w +\phi(x,t,w)s \partial_t + \xi(x,t,w) y \partial_x,
\eea
with ($x,t,w$) satisfying (\ref{e1}, \ref{e1a}, \ref{e1bb}) while ($y,s,u$) are solutions of the linear scheme prescribed by Theorem \ref{pippo}
\bea \nonumber
&&u_{m,n+1} = a_1 u_{m,n} + a_2 u_{m-1,n} + a_3 u_{m+1,n} + a_4 y_{m,n} + a_5 y_{m-1,n} + a_6 y_{m+1,n}  \\ \label{e5} && \qquad \qquad + a_7 s_{m,n} +a_8 s_{m-1,n} + a_9 s_{m+1,n} , \\ \nonumber
&&y_{m,n+1} = c_1 u_{m,n} + c_2 u_{m-1,n} + c_3 u_{m+1,n} + c_4 y_{m,n} + c_5 y_{m-1,n} + c_6 y_{m+1,n}  \\ \nonumber && \qquad \qquad +c_7 s_{m,n} + c_8 s_{m-1,n} +c_9 s_{m+1,n} , \\ \nonumber
&&s_{m,n+1} = b_1 u_{m,n} + b_2 u_{m-1,n} + b_3 u_{m+1,n} + b_4 y_{m,n} + b_5 y_{m-1,n} + b_6 y_{m+1,n}   \\ \nonumber && \qquad \qquad +b_7 s_{m,n} +b_8 s_{m-1,n} + b_9 s_{m+1,n} , 
\eea
where ($a_j,\, b_j, \, c_j$, $j=1,\cdots, 9$) are parameters at most depending on $n$ and $m$.
By a long and tedious calculation  carried out using a symbolic calculation program we get that 
\bea \label{e6}
\psi(x,t,w)&=&\psi_0(t)+\psi_1(t) x + \psi_2(t) x^2, \\ \nonumber
\phi(x,t,w) &=&\phi_0(t)+\phi_1(t) x + \phi_2(t) x^2, \\ \nonumber
\xi(x,t,w)&=&\xi_0(t)+\xi_1(t) x.
\eea
Introducing (\ref{e6}) into the determining equations for the symmetries of the discrete potential Burgers scheme (\ref{e1}, \ref{e1a}, \ref{e1bb}) we get 1672 equations for the functions ($\psi_j(t), \, \phi_j(t), \, \xi_j(t)$, $j=0,1,2$) depending on the coefficients ($a_j,\, b_j, \, c_j$, $j=1,\cdots, 9$). 168 of those equations do not depend on the coefficients ($a_j,\, b_j, \, c_j$, $j=1,\cdots, 9$) and on ($\psi_j(t+\tau), \, \phi_j(t+\tau), \, \xi_j(t+\tau)$, $j=0,1,2$); solving them imposing that $\tau \ne 0$ we get $\psi_j(t)=0$ for $j=0,1,2$, $\phi_k=0$ for $k=1,2$ and  $\xi_k=0$ for $k=0,1$. Introducing this result in the remaining 1508 equations, we get the following 9 equations
\bea \nonumber
 b_1 \phi_0(t+\tau)=b_2 \phi_0(t+\tau)=b_3 \phi_0(t+\tau)=b_4 \phi_0(t+\tau)=b_5 \phi_0(t+\tau)= \\ \nonumber =b_6 \phi_0(t+\tau)=\phi_0(t)-b_7  \phi_0(t+\tau)=b_8 \phi_0(t+\tau)=b_9 \phi_0(t+\tau)=0.
\eea
If we require $\phi_0(t)$ be not identically null, the coefficients $b_j$, $j=1,\cdots 6, 8, 9$ must be all zero and $b_7\not=0$. As a consequence $\phi_0(t)=b_7^{-n} \bar \phi$, with $ \bar \phi$ an arbitrary constant. In this case  we have a symmetry generator $\hat X =b_7^{-m} s \partial_t$ which is a consequence of the linearity of (\ref{e1bb}).
So we can conclude that the potential Burgers scheme (\ref{e1}, \ref{e1a}) is not linearizable and that the corresponding discretization of the heat equation \cite{db} is not given by a linear scheme. The linearity of the lattice equation for $t_{m,n}$ (\ref{e1bb}) is confirmed by the presence of the symmetry $\hat X =b_7^{-n} s \partial_t$.
\section*{Acknowledgements}

We thank P. Winternitz for many enlightening discussions.
LD and SC have been partly supported by the Italian Ministry of Education and Research, 
 PRIN ``Continuous and discrete nonlinear integrable evolutions: from water
waves to symplectic maps" from 2010. 
\bigskip

\bigskip


Decio Levi \\
Dipartimento di Ingegneria Elettronica (from January $1^{st}$, 2013, Dipartimento di Matematica e Fisica)\\
Universit\`a degli Studi Roma Tre and  INFN Sezione di Roma Tre \\
Via della Vasca Navale 84, 00146 Roma, Italy\\
{\it E-mail address}: {\tt levi@Roma3.infn.it}\\[0.3cm]
Christian Scimiterna \\
Dipartimento di Ingegneria Elettronica (from January $1^{st}$, 2013, Dipartimento di Matematica e Fisica)\\
Universit\`a degli Studi Roma Tre and  INFN Sezione di Roma Tre \\
Via della Vasca Navale 84, 00146 Roma, Italy\\
{\it E-mail address}: {\tt scimiterna@fis.uniroma3.it}
\label{last}
\bigskip

\end{document}